\documentclass{article}%
\usepackage{amsmath}
\usepackage{amsfonts}
\usepackage{amssymb}
\usepackage{graphicx}%
\newcommand{\bpartial}{\mathop{\partial\kern -4pt\raisebox{.8pt}{$|$}}}
\newcommand{\bra}{\mathopen{[\kern-1.6pt[}}
\newcommand{\ket}{\mathclose{]\kern-1.5pt]}}
\newcommand{\bbra}{\mathopen{[\kern-2.2pt[\kern-2.3pt[}}
\newcommand{\bket}{\mathclose{]\kern-2.1pt]\kern-2.3pt]}}

\makeindex
\begin{document}

\title{Reply to Itin, Obukhov and Hehl paper ``An Electric Charge has no Screw Sense
- A Comment on the Twist-Free Formulation of Electrodynamics by da Rocha \& Rodrigues''}
\author{Rold\~{a}o da Rocha$^{(1)}$ and Waldyr A. Rodrigues Jr.$^{(2)}$\\$^{(1)}\hspace{-0.05cm}${\footnotesize Centro de Matem\'atica, Computa\c c\~ao
e Cogni\c c\~ao}\\{\footnotesize Universidade Federal do ABC, 09210-170, Santo Andr\'e, SP,
Brazil}\\{\small \texttt{roldao.rocha@ufabc.edu.br}}\\$^{(2)}\hspace{-0.05cm}${\footnotesize Institute of Mathematics, Statistics
and Scientific Computation}\\{\footnotesize IMECC-UNICAMP CP 6065}\\{\footnotesize 13083-859 Campinas, SP, Brazil}\\{\small \texttt{walrod@ime.unicamp.br \textrm{or} walrod@mpc.com.br}}}
\date{}
\maketitle

\begin{abstract}
In this note we briefly comment a paper by Itin, Obukhov and Hehl criticising
our previous paper (\cite{roro}). We show that all remarks by our critics are
ill conceived or irrelevant to our approach and moreover we provide some
pertinent new comments to their critical paper, with the aim to clarify even
more our view on the subject.

\end{abstract}

Authors of \cite{ioh} said that it is a reaction to \cite{roro}, a paper of
ours which gives a Clifford bundle approach to classical electrodynamics. It
is our opinion that our paper deals appropriately with all their comments
(some of them unfortunately not appropriate), but here it turns out necessary
to repeat at least a \textit{crucial} remark of \cite{roro} and make some
additional few comments. The first and more important is that what we show in
our paper is that in an \textit{oriented} Lorentzian spacetime we can
formulate classical electrodynamics using only \textit{pair} form fields,
viewed as sections of an appropriate Clifford bundle (thus dispensing the use
of impair form fields\footnote{We use the the term impair forms (as originally
used by de Rham) instead the term twisted forms to avoid any sequence of words
that could seem not adequate due to one of the meanings of the word
twist in English. However, we insist here, our formulation of electrodynamics
in an \textit{oriented }spacetime does not need the use of twisted forms, but
does not claim that those objects cannot be used.}) in a coherent way using
good (but eventually not so well known) Mathematics. This is due to the fact
quoted in our paper and first spelled by de Rham \cite{derham} (an author
often quoted, often not read) that:

\textquotedblleft{\footnotesize Si la vari\'{e}t\'{e} }$V${\footnotesize est
orient\'{e}, c' est-\`{a}-dire si elle est orientable et si l"on a choisi une
orientation }$\varepsilon${\footnotesize , \`{a} toute forme impaire }$\alpha
${\footnotesize est associ\'{e}e une forme paire }$\varepsilon\alpha
${\footnotesize . Par la suite, dans le cas d'une variet\'{e} orientable, en
choissant une foi pour toutes une orientation, il serai possible d'\'{e}viter
l'emploi des formes impaires. Mais pour les vari\'{e}t\'{e}s non orientables,
ce concept est r\'{e}ellement utile et naturel}\textit{.}\textquotedblright

Using only pair forms of course, does not mean -- contrary to what our critics
think and spell -- that the resulting differential equations of our theory are
not invariant under arbitrary coordinate transformations. This is so because
all differential equations in our approach are writing intrinsically. However,
when using pair forms the sign of a charge resulting from the evaluation of
the integral of a pair current 3-form $\mathbf{J}$ depends of course, on the
handiness of the coordinate chart using for performing the evaluation. The
relevant question is: \textit{does it imply any contradiction with observed
phenomena}? As clearly shown in our paper through a very carefully analysis
using good mathematics the answer is \textit{no}. However, our critics are not
happy with our analysis and continue to insist \textit{ad nauseam} that charge
does not have a screw sense and as such the electromagnetic current must be an
impair (twisted) 3-form field $\mathbf{J}$ because they \textquotedblleft may
want to put charge on a (non-orientable) M\"{o}bius strip \ldots
\textquotedblright. Well, suppose for a while that the M\"{o}bius strip
$M\ddot{o}$ is sitting (embedded) on $\mathbb{R}^{3}$ (the rest space of an
inertial frame). To eventually calculate its charge we need to start with a
$2$-form surface charge density $\mathcal{J}$ defined on $\mathbb{R}^{3}$. Now
had our critics read our Remark 13 (see also \cite{frankel}) they could be
recalled of the fact that being $\mathcal{J}$ a pair or an impair $2$-form we
cannot define its integral over the Mobi\"{u}s strip\footnote{This could be
done only if $M\ddot{o}$ is sitting on $M\ddot{o}\times\mathbb{R}$, which is
not the case in the real physical world.}. So, we conclude that it is only in
fiction that someone can think in putting a real physical charge distribution
(made of elementary charge carriers) on a M\"{o}bius strip, and leaving aside
this physical impossibility we cannot see any necessity for the use of impair
forms. Our critics said that our statement that the Clifford bundle works only
with pair forms and could not apply to Physics if there is real need for the
use of impair forms is \textit{unsubstantied.} They justify their assertion
quoting Demers \cite{demers} which deals with a non associative `Clifford
like' algebra structure involving pair and impair forms. This structure has
nothing to do with the Clifford algebra used (as fibers) in our Clifford
bundle, which is an associative algebra, a property that makes that formalism
a very powerful computational tool. We recall also that as detailed in our
paper our formalism which writes `Maxwell equation' (no misprint here) with
pair differential forms can be split in two different ways. The first one
results in two equations using only pair forms and the second one results in
an equation using pair forms and another one using impair forms. However to do
that it is crucial to understand that there exists two different Hodge star
operators, one \textit{pair} and one \textit{impair. }They are very distinct
objects, often confused (as we explained in detail in our paper). We recall
that to have that fact in mind is important because without the
\textit{explicit} introduction of the impair Hodge dual operator the claim of
our critics (that do not even mention that object) that Maxwell equation in
the Clifford bundle splits in an equation for a pair form and one involving
impair forms is\ simply meaningless and indeed the calculation they present
(the correct ones dealing with this issue is in our paper) results in a set of
two equations involving only pair forms, contrary to their claim. Our critics
said that statement that we get from Maxwell equation the Lorentz force law is
empty because we did not define what is $F$. Well, this is simply not true. In
our approach it is clear that $F$ is taken as a physical field represented by
a $2$-form field living in Minkowski spacetime and satisfying Maxwell
equation, where a current 1-form $J$ (formed from the charged matter carriers)
acts as source of $F$. We next argued that $F$ carries energy-momentum and
that the total energy-momentum tensor of the $F$ field plus the charged matter
field is conserved. Under those well defined conditions we proved that the
coupling of $F$ with $J$ must be given by the Lorentz force law, which must
then establish the operational way in which those objects must be used when
one is doing Physics. It is in this sense that we said that such law need not
be postulated in classical electrodynamics, and we are sure that any attentive
reader of our paper will understand what we said and what we proved. A comment
is also needed, concerning the formulation of the (interesting) metric-free
approach to electrodynamics in `spacetime' defended by our critics. We leave
clear in our paper that the spacetime splitting used in their approach makes
their spacetime manifold structure closely to the Newtonian spacetime
structure. Here we remind our readers of a journey of our critics to a strange
land (which we did not visit yet and hope not to visit ever). Indeed, in
briefly reviewing the metric free approach they said that \textquotedblleft we
make minimal assumptions about spacetime, just a $4$-dimensional manifold that
we decompose into $1+3$ by means of an arbitrary normalized 4d vector
$n$.\textquotedblright\ Well, \textit{normalized} with respect to which metric
if there is no one in the metric-free approach? Finally, we recall that our
approach to the vector calculus description to Maxwell theory leaves clear
once again that we can provide a meaningful mathematical and physical
description of facts using only appropriate pair Clifford fields doing the
\textit{role} of (polar) vector fields. We say even more here, to those people
that are satisfied with the Gibbs and Heaviside approach to vector calculus
with their polar and axial vectors we leave the following issue (that
obviously did not exist in our approach). Usually $\mathbf{i}$, $\mathbf{j,k}$
are taken as a (Euclidean) orthonormal basis of polar vectors in
$\mathbb{R}^{3}$ (viewed as a vector space). Next, in vector calculus it is
introduced the vector product of two polar vectors $\mathbf{a}$ and
$\mathbf{b}$ denoted $\mathbf{a}\times\mathbf{b}$ which is said to be an
\textit{axial} vector. Next we see printed everywhere the equations
$\mathbf{i}\times\mathbf{j=k}$, $\mathbf{j}\times\mathbf{k=i}$, $\mathbf{k}%
\times\mathbf{i=j}$. Now, do $\mathbf{i}$, $\mathbf{j,k}$ become also axial by
virtue of those equations?

\end{document}